\documentclass[11pt]{article}
\usepackage[utf8]{inputenc}

\usepackage[margin=1.25in]{geometry}

\newcounter{prin}
\def\principle#1{\refstepcounter{prin}%
  {\medbreak\noindent {\bf Principle \theprin. {#1} }}%
}

\newcounter{ques}
\def\question#1{\refstepcounter{ques}%
  {\medbreak\noindent {\bf Question \theques. {#1} }}%
}

\title{Principles and Guidelines for Sharing \protect\\
Biomedical Data for Secondary Use:\protect\\
The University of Chicago Perspective}

\author{Robert L. Grossman\footnote{Corresponding author: Robert L. Grossman, University of Chicago, robert.grossman@uchicago.edu} \and Maryellen L. Giger \and Julie A. Johnson \and Jeremy D. Marks \and  Jessica P. Ridgway \and Julian Solway \and  Walter M. Stadler \protect\\ 
University of Chicago
}

\date{January 17, 2023}

\begin{document}

\maketitle

\begin{abstract} Academic medical centers are generating an increasing amount of biomedical data and there is an increasing demand for biomedical data for research purposes by research projects, research consortia, companies, and other third parties.  At the same time, as the number of patients grows and the amount of data per patient grows, there is an increasing possibility that some information about some patients may become available if the data is shared with third parties and the third parties have a data breach or violate the terms of the data use agreement.  Balancing the importance of research that may result in improved patient outcomes with the importance of protecting patient data is challenging.  The article discusses the principles, considerations about risks and mitigating risks, and guidelines used at the University of Chicago used for making decisions about sharing biomedical data with third parties.
\end{abstract}

\section{Introduction and Background}

Modern academic medical centers (AMCs) generate large amounts of clinical data from their patients. These data have allowed allowing researchers both within and outside the institution to study larger clinical populations than exist in their own institution. To protect patients and institutions, agreements between parties setting out the parameters of data to be shared have arisen, subject to legal, regulatory, and ethical review. In the last ten years, however, clinical data have been increasingly stored in comprehensive electronic relational databases, allowing study of very large populations in detail previously impossible. These electronic datasets have also been viewed by outside, for-profit entities as large corpora of organized clinical data ideal for the testing and refinement of proprietary clinical tools. On-demand platforms for data storage and analysis managed by third parties, termed cloud computing, have also become increasingly used, as have new, automated algorithms for de-identifying patient data. Accordingly, failures in the security or design of these technologies, coupled with the exponentially larger amount of data processed, raise new challenges to data security and patient privacy. In addition, the regulatory environment is becoming increasingly complex.  Finally, data sharing, even when performed in compliance with regulatory and legal requirements, risks damage to the reputations of the clinicians, researchers, and institutions involved, as patients or the public may find use of their clinical data for research, or the sharing of it with commercial entities, inappropriate. 

The above challenges are well-recognized and several approaches for making decisions about data sharing have recently been proposed; see \cite{michelson2022navigating, cole2021ten, spector2020sharing}. However, the variation of shared datasets in size and nature, and the variety of entities requesting clinical data are such that a priori rules are difficult to apply uniformly. Here, we present principles underlying clinical data sharing that that we have recognized at the University of Chicago, and guidelines that we have developed from those principles. These guidelines are designed to be utilized by data governance committees and leaders in considering the appropriateness of and mechanisms for sharing and transferring biomedical data for research purposes. 

Foremost in our considerations is the institutional obligation to patients who entrust their clinical care to an AMC. Our patients expect that their data remain private, and that any use of their data in research is for the generation of knowledge and the common good. Second, we assume that any data sharing to be evaluated will have been determined to comply with relevant regulatory and legal bodies. Finally, it is critical to remember that with all biomedical data, there is the risk that private information will be exposed.  Even when data are deidentified, “side-channel” attacks that utilize additional information can sometimes re-identify some of the individuals in the dataset \cite{erlich2014routes, gymrek2013identifying}.   This risk grows with the size of the data, so that for large de-identified datasets it is very likely that some of the individuals in it can be re-identified with sufficient side-channel information. 

Below, we describe the principles informing our approach to data sharing. Next, because risks to patient and institution are inherent in data sharing, we identify key questions to ask about the data sharing proposal. Finally, using the answers to these questions, we provide guidelines we developed for the evaluation of the acceptability of data sharing proposals.

\section{General Principles}

\principle{The patient has a right to be informed.}  Federal rules for the protection of human subjects require researchers to notify participants in some studies if commercial use of their data is possible. However, these regulations do not apply to all studies. Neither do they apply to deidentified data.  There may be other situations in which it is ethically appropriate to inform participants that data may be accessed or shared with other entities and that, in some cases, data may be used for commercial or financial benefit.
 
\principle{The institution will not sell patient data.} It is appropriate to charge for costs associated with data aggregation, data de-identification, and the efforts of faculty and staff efforts in preparing the data for sharing. However, clinical data as well as any associated clinical interpretation will not be sold to any third party.

\principle{Compliance with relevant regulatory and legal requirements.} There are a large number of local, national and international requirements for accessing, sharing and transfer of clinical data. It is the expectation that any data use or sharing is fully compliant with the relevant regulatory requirements, including whether consent needs to be obtained.

\principle{Supporting faculty and third party research must be balanced with the risks to patient privacy and to the institution's reputation.}  It is necessary to find an appropriate balance between supporting research by faculty and others benefitting the larger community, both academic research and research by commercial entities developing products, with privacy risks and institutional reputational and business risks.  These risks will depend on the size of the dataset to be shared and the nature of the data to be shared.

\principle{The institution has the ultimate responsibility for, and control of, patient data.}  As the trusted steward of its patient data, it is the AMC, rather than the individual investigator, that has final authority over the use of data derived from patient records, whether identified or de-identified. In the event of a disagreement between the AMC and any investigator, the AMC shall make any determinations regarding the proper treatment or use of any such data.

\principle{Patient right to their own clinical data.}  Patients' right to their own clinical data is recognized and will not be limited.

\section{Assessing Risks of Data Sharing to Patients and the Institution}

In this section, we identify six questions to ask that can help assess and mitigate the risks of data sharing to patients and to the institution.

\question{Who is using the data?}  We identify a hierarchy of risk in data sharing, from lowest to highest: collaborations with other research institutions, collaborations with government agencies or non-profit institutions, provision of supporting primary data to journals, sharing of data with for-profit entities.

\question{Who controls the data?} We distinguish between two settings: 1) the institution provides the data for analysis under its control; 2) the institution transfers clinical data to a third party for its use within the constraints of the data sharing agreement. The former setting is preferred. In this setting, the AMC controls who accesses the data, what data are accessed, and the access mechanism. It is only the derived data or results/findings that are then transferred to the outside entity.

\question{Are the necessary controls in place?} Unless a data management system is specifically designed for sharing data with others, it can be challenging to ensure that only the minimal required data under the relevant protocol and agreement are shared.   For data that leaves the AMC, additional scrutiny is often required, depending upon which system the data resides in, to make sure that controls are in place so that only data specified in a protocol and agreement are shared.

\question{What is the relative size of the data?} The larger the number of patients included in a dataset, the greater the chance that certain attempts to de-identify patients will succeed on at least some of the patients.  Large-scale access and larger datasets can also increase institutional reputational risks, even for participants who may have previously consented to broad reuse.

\question{With whom are the data being shared?} AMC goals include promoting greater understanding of disease, improving the delivery of clinical care, and advancing public health. These goals can be facilitated through relationships with external entities, including other AMCs, universities, not-for-profit foundations, government agencies, and commercial entities. In any data sharing arrangements, the AMC should consider the nature of the recipient, but relationships with commercial entities require special scrutiny, especially if any of the involved AMC individuals have a financial interest in the applicable entity.  

\question{What is the nature of the data and the proposed analysis?} We identify three dimensions along which the nature of the data sharing proposal for sharing may be evaluated: 1) level of identification: sharing of fully identified datasets requires greater scrutiny than does sharing of limited datasets, which, in turn, requires greater scrutiny than sharing of de-identified datasets; 2) sensitivity of the data itself: Data proposed for sharing may be highly sensitive, and require special handling and destruction. Such data may include psychotherapy notes, substance abuse data, violent trauma data, full text clinical notes, and billing data, or data from vulnerable populations; 3) focus of the proposed question for study: sharing of data for study that is exploratory in nature requires greater scrutiny than a data sharing proposal that has well defined data requested, clear analyses proposed, and defined endpoints, after which data will be destroyed.

\section{University of Chicago Medicine Guidelines for Data Sharing}
Relying on the principles and the considerations for assessing risk described above, we have developed over the course of two years the following ten guidelines: 

\begin{enumerate}

\item Data use, sharing or transfer conducted under individual informed consent is generally acceptable. To this end, novel approaches for obtaining patient consent can and should be considered including electronic consenting. It will be important to determine whether the proposed secondary use of data is compatible with the original consent or whether patient choice for secondary use of data must be obtained through a prospective, explicit consent process.

\item Data use and sharing that is summary level, fully de-identified and aggregated is generally acceptable, but the risks of reidentification should be considered. If the data contain individual-level, identifiable private information from patients or study participants, then additional scrutiny is needed.

\item	Data use, sharing or transfer of small cohorts or small slices of data for narrowly focused research projects are generally acceptable.  In contrast, the larger the cohort, the more scrutiny is required, even when, in general, data sharing falls in a setting for which a guideline has deemed generally acceptable,

\item	Data sharing with or transfer to not-for-profit entities, especially other academic institutions, and for projects conducted in the context of external peer reviewed funding is generally acceptable.

\item	Whenever possible and technically feasible, in-place queries of cohorts or slices, with appropriate protections to prevent data leakage, should be entertained. This approach is generally acceptable, especially for narrowly focused research questions. “In place” in this context can also refer to architectures that are cloud based but for which the AMC controls the data and its access.

\item	Data sharing with commercial entities with whom the PI has a conflict of interest needs to be carefully considered and may require additional COI management considerations.

\item	Data transfers of large cohorts or data slices for research projects with a broad focus and with a commercial entity are generally unacceptable.

\item	Data sharing of potentially sensitive information with commercial entities is generally unacceptable.

\item	Data sharing with and especially transfer to commercial entities for which the only or principal benefit is financial remuneration is generally unacceptable.

\item Data sharing with entities that restrict freedom of use of the data for other purposes and/or is exclusive is generally unacceptable.

\end{enumerate}

\section{Conclusions}

Modern electronic tools and large digital repositories generated in part by increasingly interconnected electronic medical records have the potential to be a rich source of information to address pressing problems and issues in disease biology and healthcare. Nevertheless, patient privacy concerns and institutional reputational and business risks become increasingly difficult to address when such data sets are widely shared, even if done so with full legal and regulatory compliance. Balancing the potential benefits of such research with these risks has thus become increasingly complex. To this end, rather than suggesting new rules or regulations, which tend to become outdated quickly as the technology changes and can impart significant administrative burden, we suggest a set of principles and guidelines to be followed by an AMC data governance structure. We anticipate that most research projects can easily be classified as acceptable or unacceptable based on these principles. We further surmise that these guidelines can be further utilized to direct discussions for those research projects for which reasonable debate on acceptability can be anticipated. 

\section{Acknowledgements}

These principles described in this paper were developed by the University of Chicago Biological Science Division Clinical Research Data Stewardship Committee.

% \bibliographystyle{plain}
% \bibliography{refs.bib}

\begin{thebibliography}{1}

\bibitem{cole2021ten}
Curtis~L Cole, Soumitra Sengupta, Sarah Rossetti, David~K Vawdrey, Michael
  Halaas, Thomas~M Maddox, Geoff Gordon, Trushna Dave, Philip~RO Payne,
  Andrew~E Williams, et~al.
\newblock Ten principles for data sharing and commercialization.
\newblock {\em Journal of the American Medical Informatics Association},
  28(3):646--649, 2021.

\bibitem{erlich2014routes}
Yaniv Erlich and Arvind Narayanan.
\newblock Routes for breaching and protecting genetic privacy.
\newblock {\em Nature Reviews Genetics}, 15(6):409--421, 2014.

\bibitem{gymrek2013identifying}
Melissa Gymrek, Amy~L McGuire, David Golan, Eran Halperin, and Yaniv Erlich.
\newblock Identifying personal genomes by surname inference.
\newblock {\em Science}, 339(6117):321--324, 2013.

\bibitem{michelson2022navigating}
Kelly~N Michelson, James~G Adams, and Joshua~MM Faber.
\newblock Navigating clinical and business ethics while sharing patient data.
\newblock {\em JAMA}, 327(11):1025--1026, 2022.

\bibitem{spector2020sharing}
Kayte Spector-Bagdady, Raymond Hutchinson, Erin~O'Brien Kaleba, and Sachin
  Kheterpal.
\newblock Sharing health data and biospecimens with industry-a
  principle-driven, practical approach.
\newblock {\em The New England journal of medicine}, 382(22):2072--2075, 2020.

\end{thebibliography}

\end{document}